\documentclass[apj]{emulateapj}

\usepackage{color}
\usepackage{CJKutf8}
\usepackage{subfigure}
\usepackage[colorlinks=true,citecolor=blue,linkcolor=blue]{hyperref}
\usepackage{amsmath}

\expandafter\def\csname 0532 \endcsname{B0532$-$67.5}
\newcommand {\snr}[1]{\csname #1 \endcsname}

\begin{document}
\begin{CJK}{UTF8}{bsmi}

\title{The Altered Chaplygin Model as a Model for Dark Energy}
\author{Yi-Syuan, Wu (吳宜軒)$^1$, Chuan-Jui Li* (李傳睿)\altaffilmark{2}, W. F. Kao (高文芳)$^1$} 
\affil{$^1$ Institute of Physics, National Yang Ming Chiao Tung University, HsinChu, 30010, Taiwan\\ homegore09@nycu.edu.tw\\
$^2$ Graduate Institute of Applied Physics, National Chengchi University, Taipei 116026, Taiwan\ \\ *cjli@nccu.edu.tw
}

%

\begin{abstract}
We present a scalar‐field formulation of the generalized Chaplygin gas (GCG) and modified Chaplygin gas (MCG) models, in which the cosmic fluid dynamics are reproduced by canonical Lagrangians with analytically derived energy density $\rho(\phi)$, pressure $p(\phi)$, and scalar potential $V(\phi)$. This framework provides a unified description of dark matter and dark energy, transitioning naturally from a matter‐dominated phase at early times to a negative‐pressure dark‐energy phase at late times. In this scalar‐field formulation, the GCG and MCG models are naturally applicable to both theoretical analyses and numerical simulations. Extending this approach, we develop a systematic method to obtain a class of integrable scalar-field cosmological models. In this study, we use this method to construct a new scalar-field  altered Chaplygin gas (ACG) model. 
To investigate the viability of Chaplygin-type models, we perform a likelihood analysis using the Pantheon+ Type Ia supernova compilation together with Cepheid-calibrated distances. We examine four models, $\Lambda$CDM, GCG, MCG, and ACG, obtaining posterior constraints on the Hubble constant $H_0$, the present‐day effective equation of state $\omega_0$, the transition redshift $z^\star$, and the cosmic age $t_0$.
With the Cepheid calibration fixing the absolute distance scale, the inferred $H_{0}$ remains nearly model-independent. 
The Chaplygin-type models predict an earlier onset of cosmic acceleration than $\Lambda$CDM and give a broader range for the inferred age of the Universe, reflecting their greater flexibility in late-time expansion histories. Among them, the ACG model provides tighter parameter constraints, while the GCG and MCG models produce broader posteriors due to parameter degeneracies.

\end{abstract}

\subjectheadings{Chaplygin gas, Dark Energy, Cosmology, $\Lambda$CDM model, Hubble constant.}

\section{ {\bf Introduction} }  \label{sec:introduction}

The accelerated expansion of the universe was first discovered through observations of distant Type Ia supernovae (SNe Ia), whose role as standardizable candles enables precise measurement of cosmological distances \citep{Riess1998, Schmidt1998, Perlmutter1999}. This discovery led to the introduction of dark energy, a component with negative pressure, to explain the observed late-time dynamics of the universe within the framework of general relativity \citep{Peebles2003, Frieman2008}.

The current standard cosmological model, $\Lambda$CDM, incorporates a cosmological constant ($\Lambda$) to account for dark energy and assumes cold dark matter (CDM) as the dominant form of non-baryonic matter. Although the model successfully explains a wide range of observations \citep{Planck2020}, including SNe Ia, the cosmic microwave background (CMB), and baryon acoustic oscillations (BAO), the fundamental origin of both $\Lambda$ and CDM remains unknown.

As an alternative to the standard $\Lambda$CDM framework, which treats dark matter and dark energy as separate components, unified dark fluid models have been proposed \citep{Bamba2012}. Among these, the generalized Chaplygin gas (GCG) model ~\citep{Chaplygin2,Kamenshchik2001,Gorini2003,MCG1,GCG} and its extension, the modified Chaplygin gas (MCG) model ~\citep{MCG1,MCG2}, describe both dark matter and dark energy as a single fluid governed by a non-standard equation of state. These models accommodate the evolution of dark matter and dark energy, enabling a smooth transition from a matter‐dominated phase at early times to a dark‐energy–dominated phase at late times. Both models share the same theoretical foundation, describe a wide range of expansion histories of the Universe, and serve as tools to test potential deviations from $\Lambda$CDM at different epochs \citep{Yang2019, Valentino2021}.

The GCG model is an effective theory with a negative pressure that decreases as the energy density decreases, similar to the typical property of dark energy ~\citep{Chaplygin2,Kamenshchik2001, Gorini2003,MCG1,GCG}. 
It is thus useful for explaining the observed acceleration of the universe produced by a cosmological constant or other forms of dark energy, and its application to the accelerating universe in the flat Friedmann-Robertson-Walker (FRW) space can be used to predict the behavior of the universe. The MCG model generalizes the GCG equation of state by adding a linear term ~\citep{MCG1,MCG2}, thereby offering greater flexibility in describing the expansion history, while remaining broadly consistent with observational constraints.

However, it is important to note that the GCG model is an effective theory with ongoing debate and research on its validity and applicability to our observed universe \citep{Carturan2003, Fabris2011}. It is noted that Chaplygin gas cosmologies face challenges at the perturbation level \citep{Sandvik2004}, where purely adiabatic perturbations can lead to oscillations and instabilities in the linear matter power spectrum \citep{Avelino2004}. More recent work \citep[e.g.,][]{Hashim2025} indicates that a sufficiently high dark sector clustering efficiency ($f \gtrsim 0.9$) could suppress these instabilities, suggesting that modified Chaplygin gas–type models may remain phenomenologically viable.
The GCG model has been constrained using various observational data, including SNe Ia \citep{Makler2003, Silva2003, Bertolami2004, Cunha2004, Zhu2004, Lu2009, Xu2012, Wang2013}, CMB \citep{Bento2003a, Amendola2003, Bean2003, Bento2003b, Lu2009, Xu2012, Wang2013}, and Hubble parameter measurements \citep{Cunha2004,Wu2007, Lu2009}, etc; similar constraints have been obtained for the MCG model \citep{Lu2008, Pourhassan2013, Paul2013, Avelino2015, Kahya2015, Li2019, Aljaf2021}. 

In cosmology, scalar fields $(\phi)$ are widely used to model the dynamics of the universe, ranging from inflationary scenarios \citep{Guth1981, Linde1982} to dark energy models \citep{Caldwell1998, Ratra1988, Gorini2003, Copeland2006}. Their flexibility in shaping the expansion history and clustering properties makes them a versatile framework for connecting high‐energy physics with cosmological observations. 

In this work, we construct scalar field realizations of the GCG and MCG models and derive the associated energy density $\rho(\phi)$, pressure $p(\phi)$, and potential $V(\phi)$ \citep{MCG1, MCG2, Wu}. Furthermore, we develop a systematic procedure to obtain a class of integrable scalar‐field cosmological models. Using this procedure, we propose a new scalar-field altered Chaplygin gas (ACG) model, which can be directly compared with other cosmological scenarios \citep{Wu}. Beyond the fluid description, the scalar field formulation of the GCG, the MCG, and the ACG models allows us to re-express their dynamics in terms of canonical Lagrangians, providing a theoretical framework for linking unified fluid models to scalar field cosmologies, and facilitating further exploration of theoretical studies, numerical simulations, and observational predictions.

To understand the late-time cosmic acceleration, 
we then carry out a full likelihood analysis of four cosmological models, the standard $\Lambda$CDM, GCG, MCG, and ACG \citep{Wu}, using the Pantheon+ SNe~Ia dataset \citep{Brout2022, Scolnic2022} together with Cepheid-calibrated distances. The analysis employed the Markov chain Monte Carlo (MCMC) method, resulting in statistically robust constraints on the cosmological parameters, including the Hubble constant $H_0$, the present-day equation-of-state parameter $\omega_0$, the transition redshift $z^{\star}$, and the age of the Universe $t_0$. Finally, we present a comparative discussion of the physical implications of these results for the nature of dark energy and the expansion history of the Universe.

Our analysis aims to evaluate the viability of the Chaplygin-type model as unified dark energy candidates ~\citep{Chaplygin2,Kamenshchik2001,MCG1,GCG,MCG2,Wu} and to assess their compatibility with current cosmological observations \citep{Riess2004, Bennett2013, Hinshaw2013, Bennett2014, Riess2021, Cunha2008, Farooq2017, Huang2020, Planck2020}. The structure of this paper is as follows: Section 2 describes the Pantheon+ dataset; Section 3 outlines the GCG, the MCG, and the new scalar-field ACG models and their scalar field formulation; Section 4 presents the inference procedure and the resulting posterior constraints for the models based on the Pantheon+ dataset; Section 5 concludes with a discussion and summary of our findings.


\section{The Pantheon+ Dataset} \label{sec:}  

The Pantheon+ compilation \citep{Brout2022, Scolnic2022} combines over 1700 spectroscopically confirmed SNe~Ia spanning the redshift range $0.001 < z < 2.26$. Pantheon+ provides improved photometric calibrations, host-galaxy redshifts, and a carefully constructed set of systematic covariance matrices that incorporate calibration, selection, and astrophysical systematics. The distance moduli are standardized using the SALT2 light-curve fitter, with nuisance parameters marginalized consistently in the construction of the dataset. This procedure ensures that robust likelihood-based inference can be performed without re-fitting individual light curves. Moreover, the inclusion of a comprehensive covariance matrix allows a statistically consistent treatment of both statistical and systematic uncertainties, making Pantheon+ one of the most powerful SNe~Ia datasets currently available for cosmological studies.

For the main cosmological inference, we employed the Pantheon+ sample to constrain four cosmological models: $\Lambda$CDM, GCG, MCG, and the new scalar-field model ACG. The analysis was carried out within a full likelihood framework using the MCMC technique. Since SNe Ia probe only relative luminosity distances \citep{Scolnic2018, Macaulay2019, Brout2022}, we explicitly incorporated the full Pantheon+ statistical plus systematic covariance matrix, along with Cepheid-calibrated distances, to ensure proper propagation of uncertainties in this framework. 
We focused on key cosmological parameters, including the Hubble constant $H_0$, the present-day equation-of-state parameter $\omega_0$, the transition redshift $z^{\star}$, and the age of the Universe $t_0$. A detailed and comparative discussion of the physical implications and observational consistency of each model is presented in the following sections.

\section{ {\bf The Scalar-Field Formulation of Cosmological Models} } \label{sec:cosmological}  

\subsection{The Modified Chaplygin Model} \label{subsec:Chaplygin}  

The generalized Chaplygin gas (GCG) model is a cosmological model representing the nature of dark energy in the universe ~\citep{Chaplygin2,Kamenshchik2001,Gorini2003,MCG1,GCG}. It is a generalization of the original Chaplygin gas model that was proposed as an effective theory to accommodate the evolution of dark matter and dark energy ~\citep{Chaplygingas}.

The equation of state of the GCG model is given by:

\begin{equation} 
p = -\frac{A}{\rho^\beta}
\end{equation}

where $p$ is the pressure, $\rho$ is the energy density, and $A$ and $\beta$ are both constants ~\citep{Chaplygin3,Chaplygin4}. 
It reduces to the original Chaplygin gas model when $\beta=1$.

The equation of state of the GCG model can be generalized to the following form:
\begin{equation}
p = \alpha \rho - \frac{A}{\rho^{\beta}}\label{eq:MCG_EoS},
\end{equation}
known as the Modified Chaplygin gas (MCG) model~\citep{MCG1,MCG2}. The MCG model has been widely studied as a candidate for unifying dark matter and dark energy into a single fluid component. 
The conservation law 
\begin{eqnarray}
{d \rho \over  \rho + p}= -d \ln a^3
\end{eqnarray}
can be integrated to give
\begin{eqnarray}
\rho = \left [  
{A \over  1 + \alpha }+ {B \over  1 + \alpha } a^{-3( 1 + \alpha )( 1 + \beta ) }
    \right ]^{1/(1+\beta ) }\label{solution_of_continuity},
\end{eqnarray}
with $B$ the integration constant. 

Instead of the fluid description
, the MCG model can be reformulated in terms of a canonical scalar field $\phi$, which reproduces the same cosmic evolution. The scalar field formulation of the MCG model admits closed-form expressions for $\rho(\phi)$, $p(\phi)$, and $V(\phi)$, offering a tractable framework for theoretical analyses and numerical implementation. For instance, the equation of state can be shown to agree with a scalar field model with the following scalar field potential 
~\citep{MCG1,MCG2}
\begin{eqnarray}
\nonumber 2V= && (1-\alpha) \left ( {A \over 1+ \alpha }\right )^{1/(1+\beta)} ~ \cosh ^{ 2/  (1+\beta)} ~{n \over 2(1+\alpha)^{1/2}} \phi
\\ && +A \left [ \left ( {A \over 1+ \alpha } \right )^{1/(1+\beta)}~\cosh ^{ 2/ (1+\beta)}~{n \over 2(1+\alpha) ^{1/2}} \phi \right ]^{-\beta}.
\end{eqnarray}
The derivation is described below.\\

{\bf Given the Lagrangian of a Scalar Field :}\\
\begin{eqnarray}
{\cal L}(\phi)=
-{1 \over 2} (\partial_\mu \phi)^2 - V(\phi).
\end{eqnarray}
It can be shown that the corresponding energy and pressure densities are given by
\begin{eqnarray}
&& 
\rho={1\over2} \dot \phi ^2 +V
\\ &&
p= {1\over2} \dot \phi ^2 -V
\end{eqnarray}
in the background FRW metric space \citep{Panotopoulos2021}.
Hence the following expression of $\dot \phi$ 
\begin{eqnarray}
\dot \phi^2= \rho +p=(1+\alpha) \rho -\frac{A}{\rho^{\beta}}\label{dot_phi_square},
\end{eqnarray}
follows directly. 
We note that the field depends only on the scale factor, $\phi = \phi(a)$. Hence the time derivative can be written as $\dot{\phi} = \phi' a H , \label{eq:phi_prime}$
where the prime denotes differentiation with respect to the scale factor $a$. Substituting $\dot \phi = \phi' aH$ into Eq. (\ref{dot_phi_square}) and we have
\begin{eqnarray}
\phi'^2 = { 
(1+\alpha) \rho -{A}/{\rho^{\beta}}
\over a^2 H^2}={ 
(1+\alpha) -{A}/{\rho^{\beta+1}}
\over a^2},
\end{eqnarray}
where we have used the Friedmann relation $H^{2} = \rho$ under the normalization $8\pi G/3 = 1$. Using the solution of the continuity equation, Eq. (\ref{solution_of_continuity}) with $B$ the integration constant, we obtain
\begin{eqnarray}
\phi' ={ [(1+ \alpha)B]^{1/2} \over a} \left ( {1 \over B+A a^n} \right )^{1/2}
\end{eqnarray}
with $n= 3(1+\alpha)(1+\beta)$.
We emphasize that in the general MCG case $n = 3(1+\alpha)(1+\beta)$, while in the special case $\alpha=0$ (the GCG limit), this reduces to $n = 3(1+\beta)$.

As a result, we can integrate $d \phi$ to obtain
\begin{eqnarray}
\phi= {\sqrt{(1+\alpha)B} \over n} \int {dx \over x\sqrt{B+Ax}},
\end{eqnarray}
with $x=a^n$.
The integral can be integrated directly to give
\begin{eqnarray}
\phi= {\sqrt{1+\alpha} \over n} \ln 
\left [
{\sqrt{B+Ax}-\sqrt{B} \over \sqrt{B+Ax} + \sqrt{B} }
\right ]
,
\end{eqnarray}
up to an integration constant $v$.
Hence we obtain the result
\begin{eqnarray}
\sqrt{B+Ax}= \sqrt{B} ~{ \cosh ~[n \phi/ 2(1+\alpha)^{1/2}]  \over \sinh ~[ n \phi / 2(1+\alpha)^{1/2}]}.
\end{eqnarray}
As a result
\begin{eqnarray}
\rho =\left ( {A \over 1+ \alpha } \right )^{1/(1+\beta)} ~ \cosh ^{ 2/  (1+\beta)} ~{n \over 2(1+\alpha)^{1/2}} \phi
.
\end{eqnarray}
The equation of motion of $\phi$ can also be shown to give
$2V= \rho -p= (1-\alpha) \rho +A \rho ^{-\beta}$.
Hence the scalar field potential is 
\begin{eqnarray}
\nonumber 2V= && (1-\alpha) \left ( {A \over 1+ \alpha }\right )^{1/(1+\beta)}\cosh ^{ 2/  (1+\beta)} ~{n \over 2(1+\alpha)^{1/2}} \phi
\\ && +A \left [ \left ( {A \over 1+ \alpha } \right )^{1/(1+\beta)}\cosh ^{ 2/  (1+\beta)}~{n \over 2(1+\alpha) ^{1/2}} \phi \right ]^{-\beta}.
\end{eqnarray}\\

{\bf Case I.} For the model $\alpha=0$.\\

This case corresponds to the standard GCG model, governed by the equation of state: $p = -\frac{A}{\rho^\beta}$ ~\citep{Chaplygin2,Kamenshchik2001,Gorini2003,MCG1,GCG}. where $\beta$ is a free parameter. The model represents a unified dark fluid scenario in which a single component accounts for both dark matter and dark energy. 

At early times (i.e., high densities), the pressure becomes negligible ($p \to 0$), resembling a pressureless matter-dominated Universe. At late times (i.e., low densities), the pressure asymptotically approaches a negative constant ($p \to -A$), effectively behaving like a cosmological constant, driving  accelerated expansion of the Universe. This model naturally captures the transition from decelerated to accelerated expansion within a single-fluid framework. 

The corresponding scalar field formulation allows for analytic expressions for $\rho(\phi)$, $p(\phi)$, and $V(\phi)$, which facilitates both theoretical analysis and numerical simulation.\\

The potential $V$ reduces to
\begin{eqnarray}
 2V=  A  ^{\frac{1}{1+\beta}}\left [\cosh ^{2/(1+\beta)}{n\over 2}\phi+\cosh^{-2\beta/(1+\beta)}{n\over2}\phi\right]
\end{eqnarray}
with $n=3(1+\beta)$.\\

{\bf Case II.} For the model $\alpha=0$, $\beta=1$. \\

This case reduces to the original (simplest) form of the Chaplygin gas model ~\citep{Chaplygingas}, characterized by the equation of state $p = -A/\rho$.
This model is fully integrable and allows simple expressions for $\rho(\phi)$ and the scalar potential $V(\phi)$. Similar to Case I, the fluid behaves like pressureless matter ($p \to 0$) at early times and smoothly transitions to a dark energy–like component ($p \to -A$) at late times. 

However, fixing $\beta = 1$ constrains the evolution more strictly and allows for analytic solutions
, making it attractive for theoretical investigation. \\

The potential $V$ reduces to
\begin{eqnarray}
 2V=  A  ^{1/2} ~ \left [\cosh  ~3\phi
 +~ \cosh ^{-1}  ~3 \phi 
\right ]
\end{eqnarray}\vspace{0.cm}

{\bf Case III.} For the model $\beta=0$.\\

This corresponds to a simplified model that deviates slightly from $\Lambda$CDM, with the equation of state $p = \alpha \rho - A$.
When $\alpha = 0$, it reduces exactly to the $\Lambda$CDM form with separate matter and cosmological constant components.
For small but nonzero $\alpha$, the model introduces mild evolution in the equation of state, enabling tests of deviations from $\Lambda$CDM.\\

Assuming that $y=\Omega_0$, $n_0 \ll 1$, for the model \citep{Bento2003a} 
\begin{eqnarray}
\rho=ya^{-(3+n_0)}+(1-y),
\end{eqnarray}
representing a small deviated model from the $\Lambda$CDM model  $\rho_{\Lambda CDM}=ya^{-3}+(1-y)$, the pressure density $p$ can be shown to be
\begin{eqnarray}
p={ n_0 \over 3} ya^{-(3+n_0)}+(y-1).
\end{eqnarray}
Here we also set $\rho_0=1$ for convenience.

It can be shown that
\begin{eqnarray}
\dot \phi ^2 =\rho+p={ y \over 3} (n_3+3) a^{-(3+n_0)}.
\end{eqnarray}
Hence the scalar potential can be integrated directly to obtain the result
\begin{eqnarray}
V={ 3-n \over 6}(1-y)\sinh^2{1\over 2}\sqrt{3(3+n_0)} \phi+(1-y).
\end{eqnarray}\\

{\bf Given a Scalar Field Potential $V(\phi)$}\\

To derive the equation of state from a given scalar potential $V(\phi)$, it can be done by observing that
\begin{eqnarray}
&& \dot \phi ^2 =\rho+p, \\
&& 2V=\rho-p .
\end{eqnarray}
Assuming $\rho=V_1(\phi)$, $p= -V_2(\phi)$, the result
\begin{eqnarray}
2V= V_1(\phi)+ V_2(\phi)
\end{eqnarray}
follows directly.
In the meantime, the continuity equation of the matter is given by 
\begin{eqnarray}
\dot \rho=-3H (\rho+p).
\end{eqnarray}
On the other hand $\dot \rho=\partial_\phi V_1 \dot \phi$, hence
\begin{eqnarray}
1=9 \rho (\rho+p) (\partial_\rho \phi)^2.
\end{eqnarray}
As a result,
\begin{eqnarray}
\rho+p = {1\over 9 \rho (\partial_\rho \phi)^2}.
\end{eqnarray}
Here the identity $\partial_\phi V_1 \partial_\rho \phi =1$ follows directly from the diffferenting of $\rho=V_1(\phi)$.
Hence
\begin{eqnarray}
&& p= {1\over 9 \rho (\partial_\rho \phi)^2}-\rho= \left [{(\partial_\phi V_1)^2\over 9 \rho^2}-1 \right ]\rho \\
&& 2V=\rho-p = \left[ 2-{(\partial_\phi V_1)^2\over 9 \rho^2} \right ] \rho .
\end{eqnarray}

{\bf Case I. $V_1= d \cosh ^e f \phi$:}\\

The pressure and scalar potential can be shown to be \begin{eqnarray}
&& p=  \left (  {e^2f^2 \over 9} -1 \right ) d \cosh ^e f\phi - {1 \over 9} d e^2f^2 \cosh ^{e-2} f\phi, \\
&& 2V=\rho-p \nonumber \\
&& \quad= \left (2-{d^2f^2\over 9 } \right ) d \cosh ^e f\phi + {1 \over 9} d e^2f^2 \cosh ^{e-2} f\phi .
\end{eqnarray}

Indeed, taking $d= {A / (1+\alpha})^{1/\beta+1}$, $e=2/(1+\beta)$ and $f=3\sqrt{1+\alpha}(1+\beta)/2$, the generalized Chaplygin gas model is redefined accordingly.\\

{\bf Case II. $V_1= d \sinh ^e f \phi$:}\\

The pressure and scalar potential can be shown to be 
\begin{eqnarray}
&& p=  \left (  {e^2f^2 \over 9} -1 \right ) d \sinh ^e f\phi + {1 \over 9} d e^2f^2 \sinh ^{e-2} f\phi, \\
&& 2V=\rho-p  \nonumber \\
&& \quad= \left (2-{d^2f^2\over 9 } \right ) d \sinh ^e f\phi -
 {1 \over 9} d e^2f^2 \sinh ^{e-2} f\phi 
\end{eqnarray}

This gives
\begin{equation}
p = \alpha \rho + \frac{A}{\rho^{\beta}},
\label{Chap2}
\end{equation}
by identifying $d= {A / (1+\alpha})^{1/\beta+1}$, $e=2/(1+\beta)$ and $f=3\sqrt{1+\alpha}(1+\beta)/2$.\\

{\bf Case III. $V_1 = d \cos^e f \phi$:}\\

The pressure and scalar potential can be shown to be
\begin{eqnarray}
&& p = \left( -\frac{e^2 f^2}{9} - 1 \right) d \cos^e(f\phi) + \frac{1}{9} d e^2 f^2 \cos^{e-2}(f\phi), \\
&& 2V = \rho - p \nonumber \\
&& \quad= \left(2 + \frac{e^2 f^2}{9} \right) d \cos^e(f\phi) - \frac{1}{9} d e^2 f^2 \cos^{e-2}(f\phi)
\end{eqnarray}

{\bf Case IV. $V_1 = d \sin^e f \phi$:}\\

The pressure and scalar potential can be shown to be
\begin{eqnarray}
&& p = \left( -\frac{e^2 f^2}{9} - 1 \right) d \sin^e(f\phi) + \frac{1}{9} d e^2 f^2 \sin^{e-2}(f\phi), \\
&& 2V = \rho - p  \nonumber \\
&& \quad= \left(2 + \frac{e^2 f^2}{9} \right) d \sin^e(f\phi) - \frac{1}{9} d e^2 f^2 \sin^{e-2}(f\phi)
\end{eqnarray}

For Case III and Case IV, the relation between the scalar field models and the MCG model follows directly.

\subsection{ The Altered Chaplygin Model } \label{subsec:ACG_Model}  

Using the Friedmann equations and the assumptions of homogeneity and isotropy in cosmology, we establish a systematic method for constructing a class of integrable scalar‐field cosmological models \citep{Wu}. Given a function $\rho(\phi)$, we can derive the corresponding pressure $p(\phi)$ and potential $V(\phi)$, as well as the equation of state $p(\rho)$. In this study, we compare the simplest integrable scalar‐field model characterized by 
\begin{gather}
Y(X)\equiv\left(\frac{dX}{d\phi}\right)^2
\end{gather}
and the energy density
\begin{gather}
    \rho(\phi)=\exp\left[k\phi^2-\lambda\right],\label{eq3.22}
\end{gather}
here $k$ and $\lambda$ are constant parameters.

Using this $\rho(\phi)$ and Eq.~(\ref{ch3eq3.3}), we can further derive the pressure $p(\phi)$:
\begin{gather}
    p(\phi)=\left(\frac{4}{9}k^2\phi^2-1\right)\exp\left[k\phi^2-\lambda\right].\label{eq3.23}
\end{gather}

The scalar potential $V(\phi)$ is then calculated using the standard relation between the potential, the scalar field, and the kinetic term $X(\phi)$, resulting in an explicit form of $V(\phi)$:
\begin{gather}
    V(\phi)=\left(1-\frac{9}{8}\alpha^2\phi^2\right)\exp\left[k\phi^2-\lambda\right].
\end{gather}

In addition to computing the energy density $\rho(\phi)$, pressure $p(\phi)$, and potential $V(\phi)$ as functions of $\phi$, we also obtain the equation of state $p(\rho)$ through the definition of equation of state parameter $\omega \equiv p/\rho$ and using Eqs.~(\ref{eq3.22}) and (\ref{eq3.23}):

\begin{gather}
    p(\rho)=\alpha\rho\ln\rho+\beta\rho.\label{eq3.26}
\end{gather}
Here we have used the fact that $\phi$ can be written as $\phi(\rho)$, $\phi^2=\frac{1}{k}\ln{\rho}+\frac{\lambda}{k}$.
Hence the pressure $p(\rho)$ can be expressed as a function of $\rho$ explicitly. It is noted that the logarithm in the ACG equation of state formally requires a dimensionless argument. To ensure dimensional
consistency, we may write instead
\[
p = \alpha \rho \ln\!\left(\frac{\rho}{\rho_\star}\right) + \beta \rho,
\]
where $\rho_\star$ is a chosen reference density. Since the additional
constant $\ln \rho_\star$ can be absorbed into a redefinition of
$\beta$, the physical content of the model is unchanged. For simplicity
and without loss of generality, we adopt $\rho_\star = 1$ throughout
this work, so that the expressions in Table \ref{fitting_table1} and subsequent equations
remain valid as written.

To simplify the expressions, we introduce two new parameters: 
$\alpha = \frac{4}{9}k$ and $\beta = \frac{4}{9}k\lambda-1$. By substituting the equation of state $p(\rho) = \alpha \rho \ln \rho + \beta \rho$ into the continuity equation, $\dot{\rho} = -3H(\rho + p)$, and integrating both sides with respect to the scale factor $a$, we arrive at the following relation:

\begin{gather}
    \int_{\rho_0}^{\rho}\frac{d\rho'}{\rho'\left[\alpha\ln\rho'+\beta+1\right]}=-3\int_{a_0}^{a}\frac{1}{a'}da',
\end{gather}

where the lower limits correspond to the present-day energy density $\rho_0$ and scale factor $a_0$, and the upper limits represent arbitrary $\rho$ and $a$. Using the relation between $a$ and redshift $z$, we further derive:
\begin{gather}    
    \frac{\alpha\ln\rho+\beta+1}{a_0^{3\alpha}\left(\alpha\ln\rho_0+\beta+1\right)}=(1+z)^{3\alpha}.\label{eq3.28}
\end{gather}
By defining $N \equiv a_0^{3\alpha}(\alpha \ln \rho_0 + \beta + 1)$ and rearranging the result, we obtain an explicit expression for $\rho(z)$ as a function of redshift $z$:

\begin{gather}
    \rho(z)=\exp\left(\frac{N}{\alpha}(1+z)^{3\alpha}-\frac{\beta+1}{\alpha}\right).
\end{gather}

Setting $\rho_0 \equiv 1$ and $a_0=1$, we find the final form of the energy density evolution $\rho(z)$:
\begin{gather}
    \rho(z)=\exp\left[\frac{\beta+1}{\alpha}\left((1+z)^{3\alpha}-1\right)\right]\label{rho_ACG}.
\end{gather}

For convenience, we summarize the resulting expressions for $\rho(\phi)$, $p(\phi)$, and $V(\phi)$. These can be written either in terms of the equation of state parameters $\alpha$ and $\beta$ (from $p = \alpha \rho \ln \rho + \beta \rho$):
\begin{equation}
\left\{
\begin{aligned}
    \rho(\phi)&=\exp\left(\frac{9\alpha\phi^2}{4}-\frac{\beta+1}{\alpha}\right)\\
    p(\phi)&=\left(\frac{9\alpha^2\phi^2}{4}-\frac{}{}1\right)\exp\left(\frac{9\alpha\phi^2}{4}-\frac{\beta+1}{\alpha}\right).\\
    V(\phi)&=\left(1-\frac{9\alpha^2\phi^2}{8}\right)\exp\left(\frac{9\alpha\phi^2}{4}-\frac{\beta+1}{\alpha}\right)
\end{aligned}
\right.\label{0.8eq0.4-5}
\end{equation}

\section{{\bf Inference from the Pantheon+} } \label{sec:}  

To evaluate the viability of the Chaplygin-type models \citep{Chaplygin2,Kamenshchik2001,MCG1,GCG,MCG2,Wu}, we perform a full likelihood analysis using the Pantheon+ SNe~Ia dataset ($0.001 < z < 2.26$; \citealt{Brout2022, Scolnic2022}). In our analysis, we employ the MCMC technique to obtain posterior constraints on the cosmological parameters, incorporating the full statistical and systematic covariance matrix together with Cepheid-calibrated distances. The analysis considers four cosmological models: the standard $\Lambda$CDM, GCG, MCG, and ACG \citep{Wu}, focusing on four key parameters: the present-day Hubble constant $H_0$, the equation-of-state parameter $\omega_0$, the transition redshift $z^{\star}$, and the age of the Universe $t_0$. This framework allows us to examine the ability of each model to reproduce current cosmological observations and to assess their respective dynamical implications.

\begin{table}[h!]
\caption{Fitting Functions and Parameters of Models}
\hspace{-0.4cm}
\begin{tabular}{c p{0.8\linewidth} c c }
\hline\hline
Model & Equation of State &  &  \\  [0.5ex] 
\hline
$\Lambda$CDM  & $p=-\rho$ &   & \\
GCG            &  $p=-A\rho^{-\beta}$ &   & \\
MCG            & $p=\alpha\rho-A\rho^{-\beta}$ &   & \\
ACG           & $p=\alpha\rho\ln\rho+\beta\rho$ &   & \\
[1ex]
\hline\hline
Model & Energy Density &  &  \\  [0.5ex] 
\hline
$\Lambda$CDM  & $\rho(z)=\rho_{M0}(1+z)^3+(1-\rho_{M0}) $ &   & \\
GCG            & $\rho(z)=\left[A^{\star}+(1-A^{\star})(1+z)^{3(1+\beta)}\right]^{\frac{1}{1+\beta}}$  &   & \\
MCG            & $\rho(z)=\left[A_0+(1-A_0)(1+z)^{3(1+\alpha)(1+\beta)}\right]^{\frac{1}{1+\beta}}$ &   & \\
ACG           & $\rho(z)=\exp\left[\frac{\beta+1}{\alpha}\left((1+z)^{3\alpha}-1\frac{}{}\right)\right]$ &   & \\
[1ex]
\hline\hline
Model & Fitting Function &  &  \\  [0.5ex] 
\hline
$\Lambda$CDM  & $ F(z)=\int^z_0\frac{dz'}{\sqrt{\rho_{M0}(1+z')^3+(1-\rho_{M0})}} $ &   & \\
GCG            &  $ F(z)=\int^{z}_0\frac{dz'}{\sqrt{\left[A^{\star}+(1-A^{\star})(1+z')^{3(1+\beta)}\right]^{\frac{1}{1+\beta}}}} $ &   & \\
MCG            & $F(z)=\int^{z}_0\frac{dz'}{\sqrt{\left[A_0+(1-A_0)(1+z')^{3(1+\alpha)(1+\beta)}\right]^{\frac{1}{1+\beta}}}} $ &   & \\
ACG           & $  F(z)=\int^{z}_0\exp\left[\frac{\beta+1}{2\alpha}\left(\frac{}{}1-(1+z')^{3\alpha}\right)\right]dz' $ &   & \\
[1ex]
\hline\hline
Model & Equation of State Parameter &  &  \\  [0.5ex] 
\hline
$\Lambda$CDM  & $\omega(z)=\frac{\rho_{M0}(1+z)^3}{\rho_{M0}(1+z)^3+(1-\rho_{M0})}-1$ &   & \\
GCG            & $\omega(z)=-\frac{A^{\star}}{A^{\star}+(1-A^{\star})(1+z)^{3(1+\beta)}}$ &   & \\
MCG            & $\omega(z)=\alpha-\frac{A_0(1+\alpha)}{A_0+(1-A_0)(1+z)^{3(1+\alpha)(1+\beta)}}$ &   & \\
ACG           & $\omega(z)=(1+\beta)(1+z)^{3\alpha}-1$ &   & \\
[1ex]
\hline\hline
Model & Transition Redshift &  &  \\  [0.5ex] 
\hline
$\Lambda$CDM  & $z^{\star}=\sqrt[3]{\frac{2(1-\rho_{M0})}{\rho_{M0}}}-1$ &   & \\
GCG            & $z^{\star}=\left(\frac{2A^{\star}}{1-A^{\star}}\right)^{\frac{1}{3(1+\beta)}}-1$ &   & \\
MCG            & $z^{\star}=\left[\frac{2A_0}{(1+3\alpha)(1-A_0)}\right]^{\frac{1}{3(1+\alpha)(1+\beta)}}-1$ &   & \\
ACG           & $z^{\star}=\left[\frac{2}{3(\beta+1)}\right]^{\frac{1}{3\alpha}}-1$ &   & \\ [1ex]
\hline\hline
Model & Age of the Universe &  &  \\  [0.5ex] 
\hline
$\Lambda$CDM  & $t_0\equiv\frac{1}{H_0}\int^{\infty}_0\frac{1}{(1+z')\sqrt{\rho_{M0}(1+z')^3+(1-\rho_{M0})}}dz'$ &   & \\
GCG            &  $t_0\equiv\frac{1}{H_0}\int^{\infty}_{0}\frac{1}{(1+z')\left[A^{\star}+(1-A^{\star})(1+z')^{3(1+\beta)}\frac{}{}\right]^{\frac{1}{2(1+\beta)}}}dz'$ &   & \\
MCG            & $t_0\equiv\frac{1}{H_0}\int^{\infty}_{0}\frac{1}{(1+z')\left[A_0+(1-A_0)(1+z')^{3(1+\alpha)(1+\beta)}\frac{}{}\right]^{\frac{1}{2(1+\beta)}}}dz'$ &   & \\
ACG           & $t_0\equiv\frac{1}{H_0}\int^{\infty}_{0}\frac{1}{(1+z')\exp\left[\frac{\beta+1}{2\alpha}\left((1+z')^{3\alpha}-1\right)\right]}dz'$ &   & \\[1ex]
\hline
\end{tabular}
\tablenotetext{1}{To simplify the fitting procedure, we set $\rho_0 \equiv 1$ and correspondingly rescaled the relevant functions and parameters \citep{Hogg1999}. For the GCG model, this gives $A = \rho_0 A^{\star}$. For the MCG model, we adopt $A = A_0 \rho_0 (1 + \alpha)$. Here, $\rho_{M0}$ denotes the present-day matter energy density of the Universe, and the reference density is given by $\rho_0 = 1.8788\,H_0^2 \times 10^{-30}\ \mathrm{kg\ m^{-3}}$. 
}
\label{fitting_table1}
\end{table}

\subsection{Cosmological Functions and Parameters}

In a spatially flat universe ($k = 0$), the luminosity distance $d_L$ is related to the energy density $\rho(z)$ through the following definitions~\citep{EdwardWKolb, WCosmology}:
    $d_L=10^{1+\mu/5}$, 

where $\mu$ is the distance modulus, $c$ is the speed of light, and $H_0$ is the present-day Hubble constant. The luminosity distance $d_L$ is expressed in parsecs. 
We then introduce the dimensionless luminosity distance 
\begin{gather}
D_L(z) \equiv H_0 d_L(z)/c = (1+z) \int^{z}_0 dz'/E(z'), 
\end{gather}
where $E(z) \equiv H(z)/H_0 = \sqrt{\rho(z)}$ with the normalization $\rho_0 = 1$ \citep{Hogg1999}.
This gives $\mu = 5 \log_{10} [(c/H_0) D_L /pc] - 5$ and a fitting function $F(z)$ 
, which serves as the observational input for parameter estimation ~\citep{Chang, Tsai}:
\begin{gather}
    F(z)=\int^{z}_0\frac{1}{E(z')}dz',\label{F_z}
\end{gather}

Based on the energy conservation equation in an expanding universe,
$\dot{\rho} = -3H(1 + \omega)\rho$, and using the relation $\frac{d(1+z)}{da} = -(1+z)^2$, we can derive the redshift-dependent total equation of state parameter as
\begin{equation}
\omega(z) = \frac{1+z}{3 \rho}\frac{d\rho}{dz}-1.\label{omega_z}
\end{equation}
This expression allows us to compute the effective $\omega(z)$ directly from the redshift evolution of the energy density $\rho(z)$.

In particular, the Universe transitions from decelerating to accelerating expansion when the equation of state crosses $\omega = -1/3$, which we define as the critical value $\omega^{\star}$. The corresponding redshift at which this occurs is the transition redshift $z^{\star}$.
\begin{equation}
z^{\star} = \frac{2 \rho(z^\star)}{ \left. d\rho/dz \right|_{z = z^\star} } - 1.\label{z_star}
\end{equation}

Since we normalize the energy density to the present-day value $\rho_0 = 3H_0^2 / (8\pi G)$, we start from the Friedmann equation $H = \dot{a}/a = H_0 \sqrt{\rho(z)}$ and use the relation between redshift and scale factor, $1+z = 1/a$. 
By separating variables and integrating with respect to $z$, we obtain the expression for the cosmic age as a function of redshift $t(z)$:
\begin{gather}
    t(z)\equiv\frac{1}{H_0}\int^{\infty}_{z}\frac{1}{(1+z')E(z')}dz'\label{t_0},
\end{gather}
with the present-day age of the Universe defined as $t_0 \equiv t(z=0)$. Throughout this work, $t_0$ is expressed in gigayears (Gyr).

By substituting the energy densities of the $\Lambda$CDM, GCG, MCG, and ACG models into equations (\ref{F_z}-\ref{t_0}),
we can compute the fitting function $F(z)$, derive the Hubble constant H$_0$, the redshift-dependent equation of state $\omega(z)$, the transition redshift $z^{\star}$, and the age of the Universe $t_0$ for each model. For practical fitting, we recast the model‐dependent functions and parameters into simplified forms, as summarized in Table~\ref{fitting_table1}. Such normalization is a standard practice in cosmological parameter estimation, ensuring dimensionless quantities and simplifying numerical analysis. These results are constrained 
using the Pantheon+ compilation \citep{Brout2022, Scolnic2022} for the full set of four parameters. The outcomes are then compared with current observational constraints.

\subsection{ Likelihood-Based Inference } \label{subsec:}  

For the main cosmological inference, we employed a likelihood-based framework on the Pantheon+ dataset \citep{Brout2022, Scolnic2022}, consisting of over 1700 SNe~Ia spanning $0.001 < z < 2.26$. The data vector of observed distance moduli $\mu_{\rm obs}$, calibrated using Cepheid host galaxies, was compared against the model predictions $\mu_{\rm th}(z;\theta)$ using the full statistical plus systematic covariance matrix provided by Pantheon+. The log-likelihood was taken as the standard multivariate Gaussian form:
\begin{equation}
\begin{aligned}
\ln \mathcal{L} = &-\tfrac{1}{2} \big[ (\mu_{\rm obs}-\mu_{\rm th})^T C^{-1} (\mu_{\rm obs}-\mu_{\rm th}) \\
&+ \ln \det C + N \ln (2\pi) \big],
\end{aligned}
\end{equation}
where $\theta$ is the set of cosmological parameters for a given model, and $C$ denotes the Pantheon+ statistical plus systematic covariance matrix. Cosmological posteriors were obtained using the MCMC algorithm, ensuring proper marginalization over all nuisance directions. Within this framework, we analyzed four representative dark-energy models, $\Lambda$CDM, GCG, MCG, and ACG, extracting constraints on the Hubble constant $H_0$, the present effective equation-of-state parameter $\omega_0$, the transition redshift $z^\star$, and the cosmic age $t_0$. This approach provides statistically robust confidence intervals that properly propagate both statistical and systematic uncertainties.

\begin{table*}
\caption{ Posterior Constraints on Cosmological Parameters Using the Pantheon+ Compilation$^a$ }
\begin{tabular}{c c c c c l l l} 
\hline\hline
\rule{0pt}{9pt} Pantheon+ & SNe Ia + Cepheid & \multicolumn{2}{l}{Marginalization fits}  &  &  &  &  \\[2pt]
\hline
\rule{0pt}{9pt} Model & $H_0$ (km s$^{-1}$ Mpc$^{-1}$) & $\omega_0$ & $z^{\star}$ & $t_0$ (Gyr) & Model Parameters &  &  \\[2pt] 
\hline 
\rule{0pt}{11pt} $\Lambda$CDM & $72.83_{-0.23}^{+0.23}$ & $-0.64_{-0.02}^{+0.02}$ & $0.52_{-0.04}^{+0.04}$ & $12.28_{-0.15}^{+0.16}$  & $\Omega_{m, 0}=0.36_{-0.02}^{+0.02}$ & - & -\\
\rule{0pt}{11pt} GCG & $72.59_{-0.25}^{+0.26}$  & $-0.58_{-0.04}^{+0.03}$ & $0.86_{-0.25}^{+0.47}$ & $13.21_{-0.62}^{+0.73}$  & $A^*=0.58_{-0.03}^{+0.04}$ & - & $\beta=-0.46_{-0.18}^{+0.25}$\\
\rule{0pt}{11pt} MCG & $72.58_{-0.26}^{+0.25}$  & $-0.58_{-0.04}^{+0.03}$ & $0.92_{-0.28}^{+0.52}$ & $13.57_{-1.01}^{+0.79}$ & $A_0=0.56_{-0.11}^{+0.12}$ & $\alpha=-0.05_{-0.20}^{+0.39}$ & $\beta=-0.40_{-0.32}^{+0.74}$ \\
\rule{0pt}{11pt} ACG & $72.55_{-0.26}^{+0.26}$  & $-0.57_{-0.03}^{+0.03}$ & $0.85_{-0.21}^{+0.45}$ & $12.09_{-1.00}^{+1.18}$  & - & $\alpha=0.23_{-0.08}^{+0.10}$ & $\beta=-0.57_{-0.03}^{+0.03}$  \\[2pt]
\hline
\end{tabular}
\tablenotetext{1}{Total 1700+ SNe~Ia with $0.001 < z < 2.26$ \citep{Brout2022, Scolnic2022}.}
\label{fitting_table2}
\end{table*}

\subsection{Posterior Constraints from Pantheon+} \label{subsec:}  

\begin{figure*}[]
    \includegraphics[width=\textwidth]{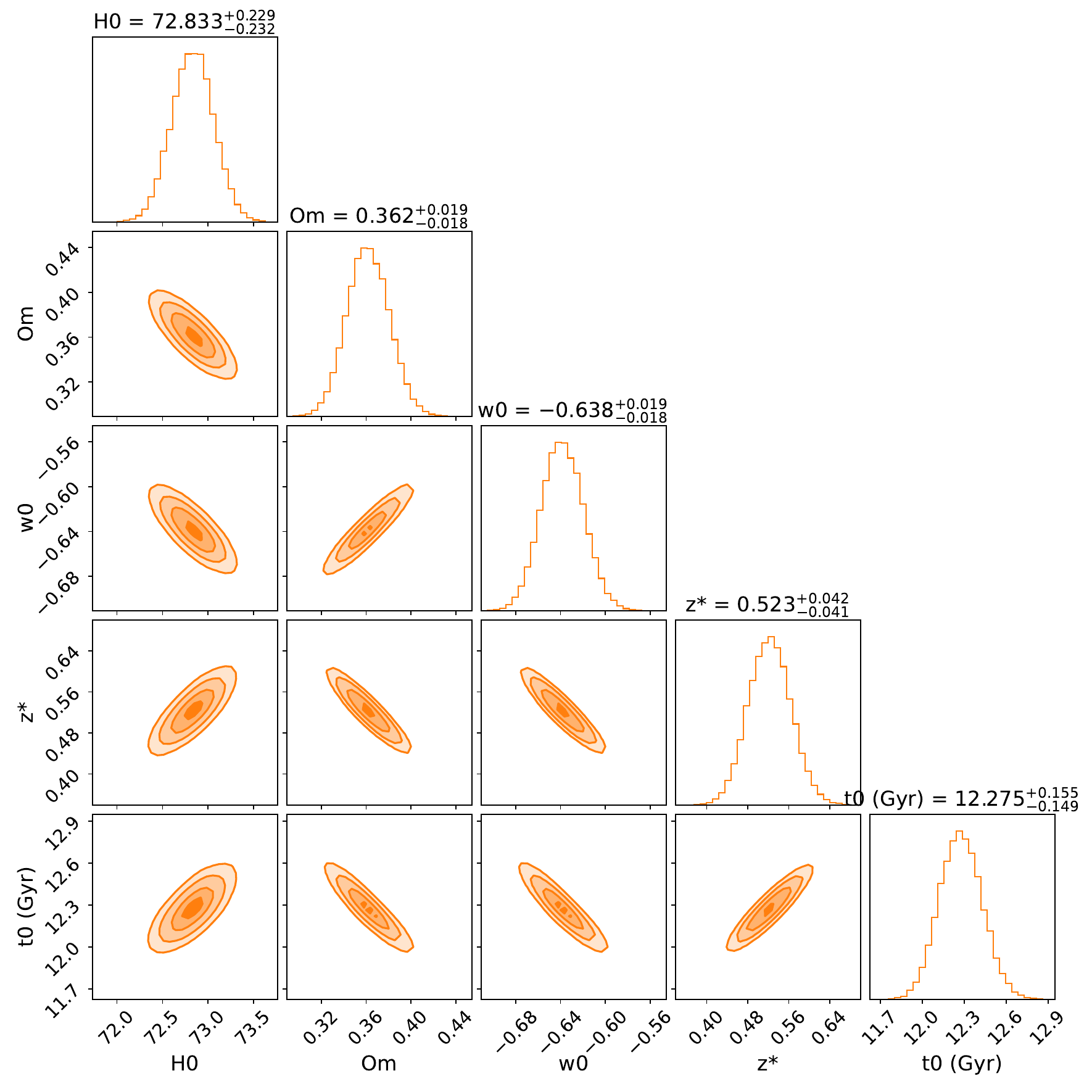}
    \caption{Posterior constraints from the Pantheon+ dataset \citep{Brout2022}, including SNe~Ia in the redshift range $0.001 < z < 2.26$ with Cepheid-calibrated distances. The marginalized posterior estimates of the Hubble constant $H_0$, the present-day matter density parameter, the present-day effective equation-of-state parameter $\omega_0$, the transition redshift $z^{\star}$, and the age of the Universe $t_0$ for the $\Lambda$CDM model are summarized in Table~\ref{fitting_table2}.}
    \label{MCMC_LCDM_Color}
\end{figure*}

\begin{figure*}[]
    \includegraphics[width=\textwidth]{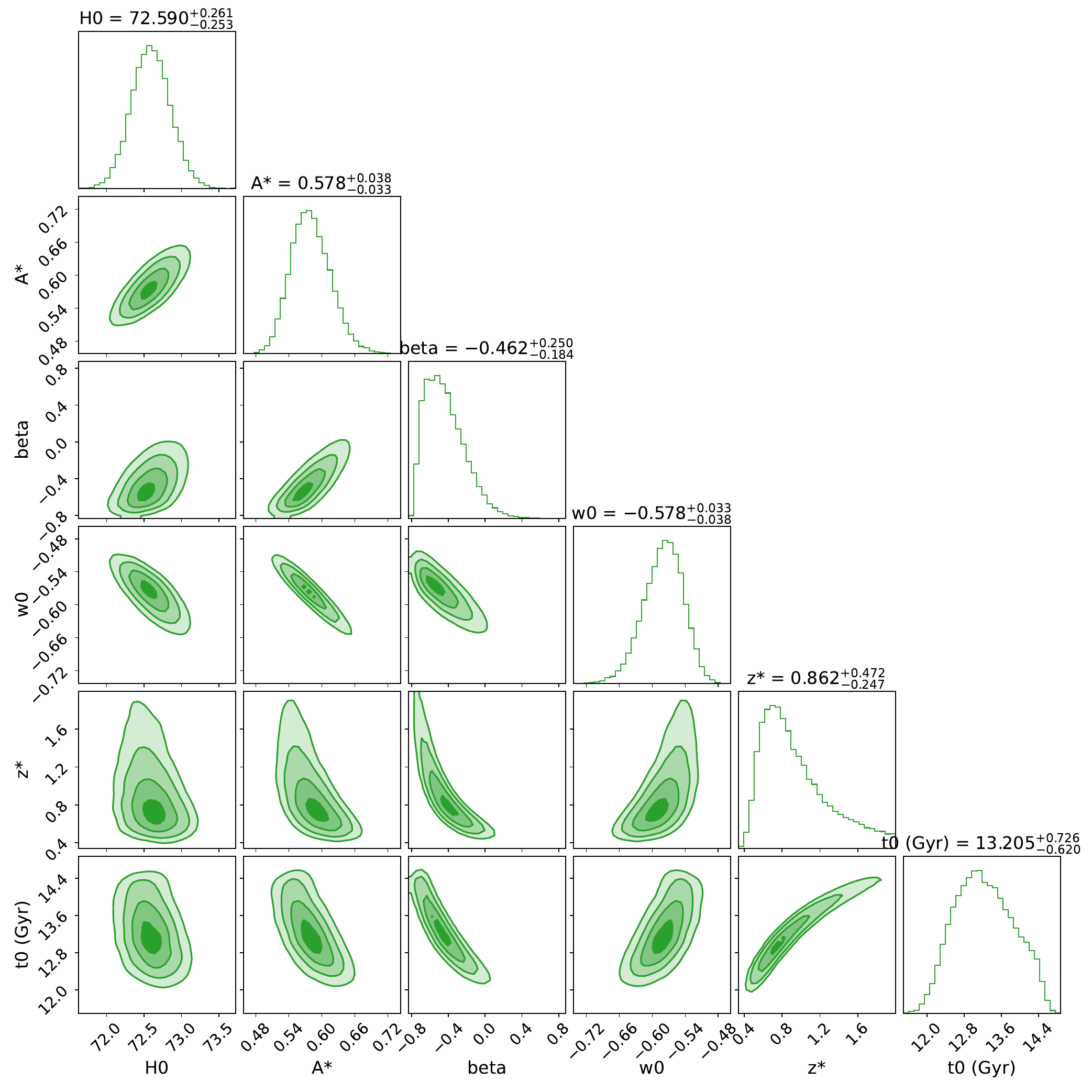}
    \caption{Same as Figure~\ref{MCMC_LCDM_Color}, but for the GCG model. The corresponding parameters are also summarized in Table~\ref{fitting_table2}.}
    \label{MCMC_GCG_Color}
\end{figure*}

\begin{figure*}[]
    \includegraphics[width=\textwidth]{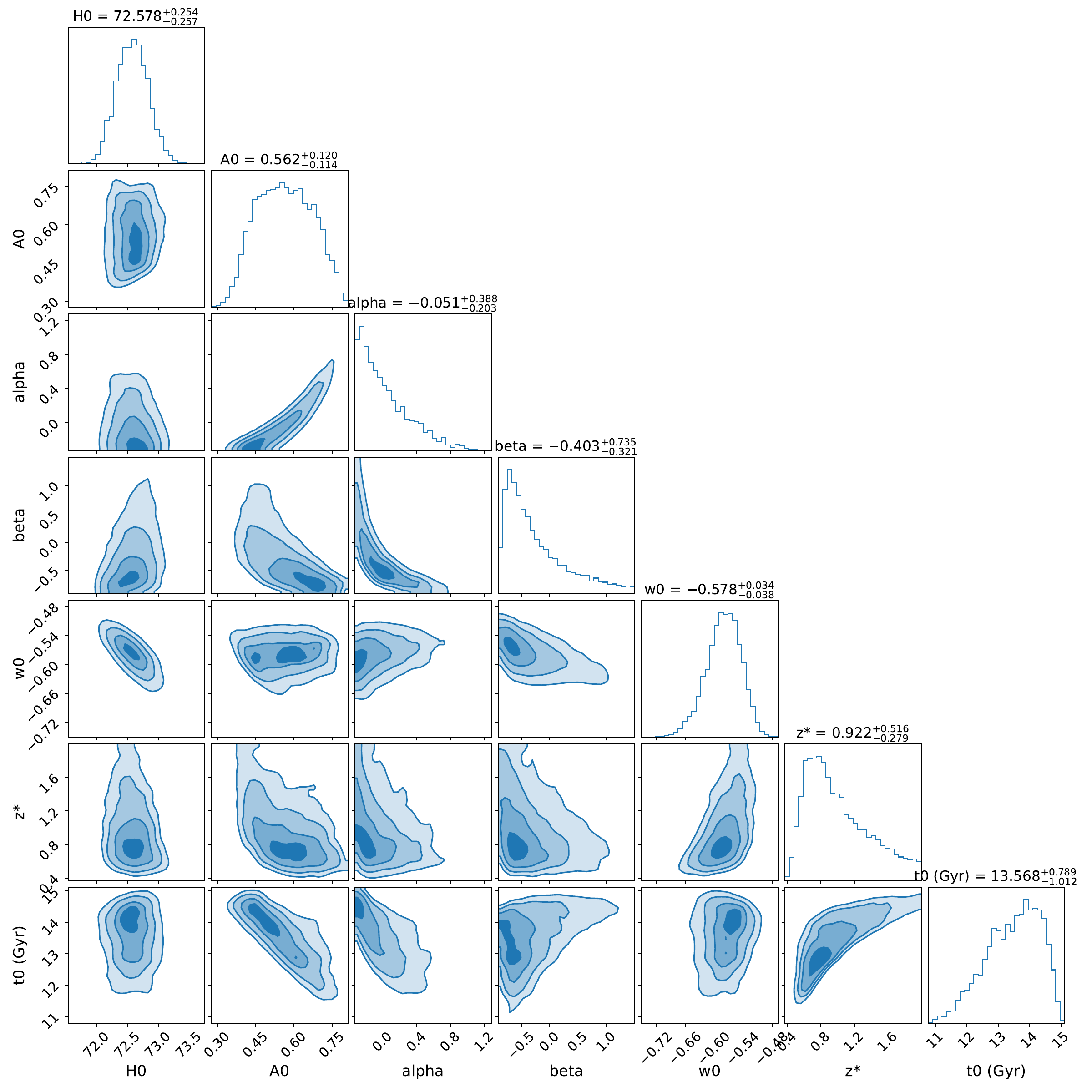}
    \caption{Same as Figure~\ref{MCMC_LCDM_Color}, but for the MCG model. The corresponding parameters are also summarized in Table~\ref{fitting_table2}.}
    \label{MCMC_MCG_Color}
\end{figure*}

\begin{figure*}[]
    \includegraphics[width=\textwidth]{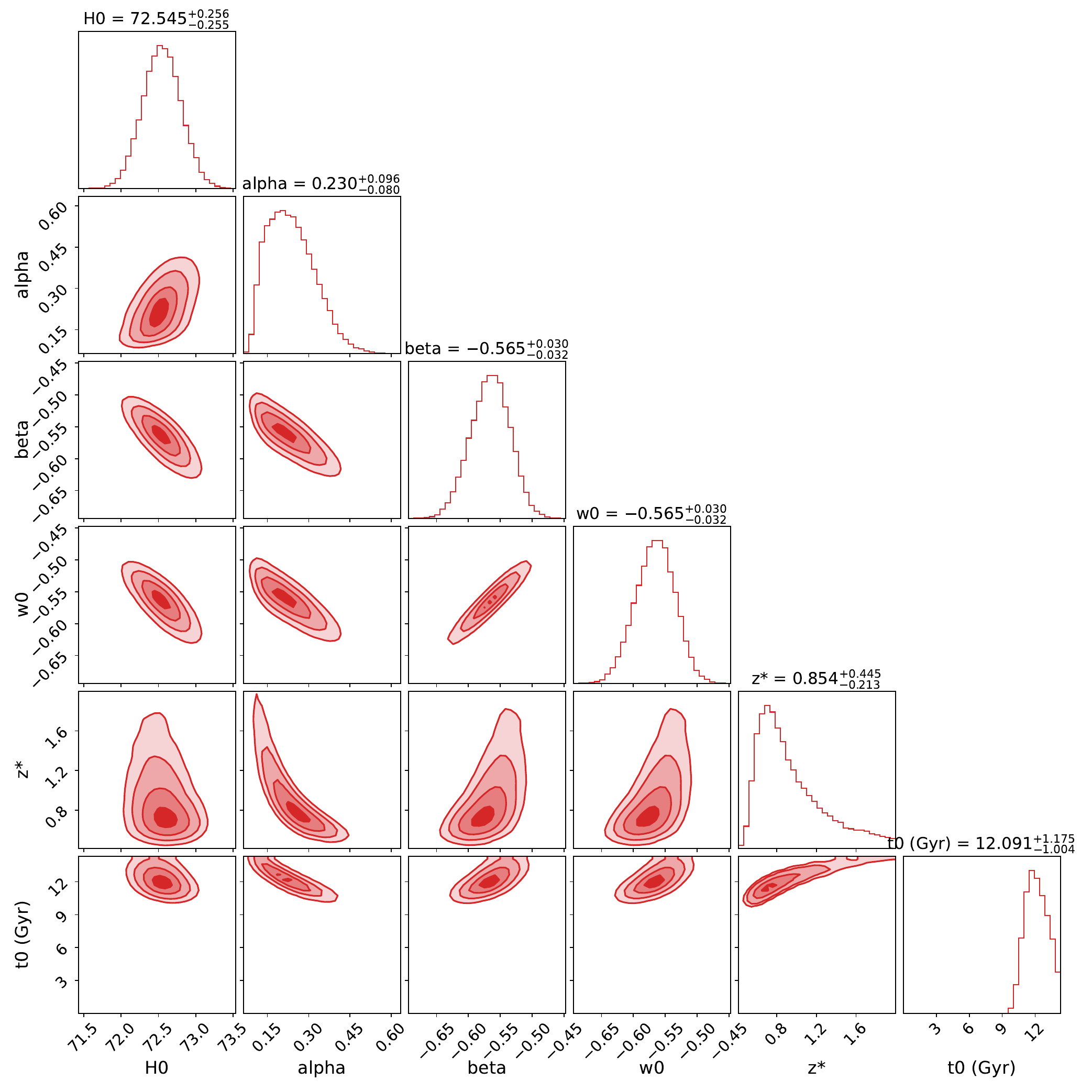}
    \caption{Same as Figure~\ref{MCMC_LCDM_Color}, but for the ACG model. The corresponding parameters are also summarized in Table~\ref{fitting_table2}.}
    \label{MCMC_ACG_Color}
\end{figure*}

For our main cosmological inference, we employed the MCMC technique within a full likelihood framework. Using the Pantheon+ dataset \citep{Brout2022, Scolnic2022}, which incorporates Cepheid-calibrated SNe Ia's distances and the full statistical plus systematic covariance matrix, we obtained marginalized posterior constraints on the cosmological parameters of interest. The numerical results of these parameters are summarized in Table~\ref{fitting_table2}, while the corresponding marginalized posterior distributions of parameters are displayed in Figures~\ref{MCMC_LCDM_Color}--\ref{MCMC_ACG_Color}. In these figures, the posterior distributions of the Hubble constant $H_0$, the present-day effective equation-of-state parameter $\omega_0$, the transition redshift $z^{\star}$, and the cosmic age $t_0$ are explicitly shown, directly corresponding to the numerical values reported in the table.


Across all four models, the inferred Hubble constant is remarkably consistent, with values clustering tightly around $H_0 \simeq 72.55{-}72.83$ km s$^{-1}$ Mpc$^{-1}$. This agreement reflects the strong Cepheid calibration built into Pantheon+SH0ES, which effectively anchors the absolute distance scale \citep{Riess2021, Riess2022}.
Our constraints align well with the SH0ES determination of $H_0 = 73.04 \pm 1.04$ km s$^{-1}$ Mpc$^{-1}$ \citep{Riess2022}, the Miras-calibrated SN Ia result of $H_0 = 73.3 \pm 4.0$ km s$^{-1}$ Mpc$^{-1}$ \citep{Huang2020}, and the time-delay lensing measurement from the H0LiCOW collaboration of $H_0 = 73.3^{+1.7}_{-1.8}$ km s$^{-1}$ Mpc$^{-1}$ \citep{Wong2020}, while remaining in notable tension with the lower Planck CMB-inferred value of $67.4 \pm 0.5$ km s$^{-1}$ Mpc$^{-1}$ \citep{Planck2020}. For comparison, the WMAP nine-year CMB-only analysis gives $H_0 = 70.0 \pm 2.2$ km s$^{-1}$ Mpc$^{-1}$, while the combination of WMAP nine-year with 
CMB, BAO, and a local $H_0$ prior predicts $H_0 = 69.32 \pm 0.80$ km s$^{-1}$ Mpc$^{-1}$ \citep{Bennett2013, Hinshaw2013, Bennett2014}. 
Previous work has demonstrated that the local distance-ladder determination of $H_{0}$ is largely insensitive to the assumed dark energy model once the Cepheid calibration is included, with inter-model variations of $0.47\ \mathrm{km\ s^{-1}\ Mpc^{-1}}$, i.e., a 0.6\% shift in $H_{0}$, which is significantly smaller than the observed tension \citep{Dhawan2020}. This reflects the fact that SNe Ia provide only relative distances, and the absolute scale is anchored by the Cepheid calibration \citep{Scolnic2018, Macaulay2019, Brout2022}.

For $\Lambda$CDM, the marginalized matter density is 
$\Omega_{m,0} = 0.36^{+0.02}_{-0.02}$, generally consistent with SNe-only analyses such as Pantheon+ 
($\Omega_{m,0} = 0.334 \pm 0.018$; \citealt{Brout2022}), and slightly higher than the Planck 2018 baseline of $\Omega_{m,0} = 0.315 \pm 0.007$ \citep{Planck2020}. Our corresponding present-day effective equation of state is $w_{0} = -\Omega_{\Lambda,0} = -(1-\Omega_{m,0})$, giving $w_{0} = -0.64^{+0.02}_{-0.02}$. For comparison, Pantheon+ implies $w_{0} \approx -0.666$ \citep{Brout2022}, while Planck 2018 gives $w_{0} \approx -0.685$ \citep{Planck2020}. In contrast, the Chaplygin-type models do not treat $\Omega_{m, 0}$ as a direct free parameter. Instead, they introduce phenomenological parameters ($A^*$ or $A_0$, $\alpha$, $\beta$) that characterize the relative contributions of the matter-like and negative-pressure components \citep{Chaplygin2,Kamenshchik2001,MCG1,GCG,MCG2,Wu}.

For the transition redshift, the $\Lambda$CDM model provides $z^{\star} = 0.52_{-0.04}^{+0.04}$. In contrast, Chaplygin-type models give higher values, reaching up to $z^{\star} = 0.92^{+0.52}_{-0.28}$ for MCG. This behavior arises from the unified dark fluid description, in which a single component contains both matter-like and negative-pressure terms, thereby allowing more flexible dynamics than $\Lambda$CDM \citep{Chaplygin2, Kamenshchik2001, Gorini2003, MCG1, GCG, MCG2}. As the Universe expands, the negative-pressure contribution in these models declines more slowly than in $\Lambda$CDM and therefore becomes dynamically important at an earlier epoch, advancing the onset of acceleration. 
While the uncertainties for Chaplygin-type models are relatively large, the results remain broadly consistent with independent determinations of $z^\star$ \citep{Riess2004, Wu2007b, Cunha2008, Lu2009, Farooq2017}, suggesting that these models remain viable descriptions of the expansion history.

The inferred cosmic ages span from $t_0 \approx 12.09_{-1.00}^{+1.18}$ Gyr (ACG), up to $\approx 13.57_{-1.01}^{+0.79}$ Gyr (MCG), while the model-dependent spread highlights the sensitivity of $t_0$ to the chosen dark-energy parameterization. The $\Lambda$CDM and ACG give $t_0 \approx 12.28_{-0.15}^{+0.16}$ Gyr and $12.09_{-1.00}^{+1.18}$, respectively.
Both the $\Lambda$CDM and ACG estimates are more younger than the CMB-based values, $13.80 \pm 0.02$ Gyr from Planck \citep{Planck2020} and $13.772 \pm 0.059$ Gyr from the WMAP9 analysis combined with external datasets \citep{Bennett2013, Hinshaw2013, Bennett2014}, with the difference driven by the higher $H_0$ enforced by the Cepheid-calibrated Pantheon+ distances. In these models, the expansion history $H(z)$ follows a more constrained form, so that a higher $H_0$ directly reduces the integrated age.
Their relatively younger ages nevertheless remain compatible with the lower bound from the old Galactic star HD~140283, the so-called “Methuselah star” \citep{Bond2013, Tang2021}. 
By contrast, the GCG and MCG models introduce additional degrees of freedom in their equations of state, allowing $H(z)$ to evolve more slowly at intermediate redshifts. This added flexibility mitigates the impact of a high $H_0$, resulting in longer cosmic ages of $13.21_{-0.62}^{+0.73}$ and $13.57_{-1.01}^{+0.79}$ Gyr, respectively.
The upper bound of the cosmic age recovered here, $ 13.57_{-1.01}^{+0.79}$ Gyr, is also consistent with independent age determinations from the oldest globular clusters. $\approx$ 13.8 Gyr \citep{Valcin2021, Ying2023}.


Regarding the model parameters, for the GCG we obtain $A^\star = 0.58^{+0.04}_{-0.03}$ and $\beta = -0.46^{+0.25}_{-0.18}$, generally
consistent with earlier constraints that favor small or mildly negative values of $\beta$ \citep{Lu2009}. The MCG parameters are less tightly constrained, with $A_0 = 0.56^{+0.12}_{-0.11}$, $\alpha = -0.05^{+0.39}_{-0.20}$, and 
$\beta = -0.40^{+0.74}_{-0.32}$, reflecting parameter degeneracies noted in prior work and underscoring the need for complementary probes such as BAO and CMB \citep{MCG1, Lu2008, Paul2013}. For the ACG model, we find $\alpha = 0.23^{+0.10}_{-0.08}$ and $\beta = -0.57^{+0.03}_{-0.03}$, with much tighter bounds than those of GCG or MCG. This suggests that the ACG parameters can be better constrained by SNe~Ia data calibrated with Cepheids over the range $0.001 < z < 2.26$, while remaining consistent with independent observational estimates. 

Overall, the consistency of these Chaplygin-type models with the Pantheon+ likelihood analysis indicates that they remain phenomenologically viable, offering an alternative interpretation of cosmic acceleration. Our posterior estimates are directly comparable to independent results obtained from diverse data sets and modeling frameworks \citep{Gonzalez-Diaz2003, Lu2008, Lu2009, Xu2012, Hernandez-Almada2019, Dainotti2021, Zheng2022}. Future incorporation of BAO, CMB, and other complementary probes will be crucial to strengthen constraints and clarify their physical implications.

\section{ {\bf Discussions and Summary} } \label{sec:discussions}  

In this study, we derived the scalar‐field formulation corresponding to the MCG model ~\citep{Chaplygin3,Chaplygin4}. By rewriting the MCG equation of state in terms of the Lagrangian of a canonical scalar field \citep{Panotopoulos2021}, we obtained closed‐form expressions for the energy density $\rho(\phi)$, pressure $p(\phi)$, and potential $V(\phi)$. This formulation offers a more intuitive dynamical interpretation and facilitates its application in both theoretical analyses and numerical simulations \citep{Wu}. In this scalar‐field formulation, the MCG model can be reduced to the standard GCG model \citep{Chaplygin2,Kamenshchik2001,Gorini2003,MCG1,GCG} as $\alpha=0$, to the original form of the Chaplygin gas model \citep{Chaplygingas} as $\alpha=0$ and $\beta=1$, and to a simplified model that slightly deviates from the $\Lambda$CDM model as $\beta=0$. In addition, we can also derive the equation of state if a scalar field potential $V(\phi)$ is given.

The scalar‐field form reveals that, in the early universe, the MCG can be approximated as pressureless dark matter–dominated, while at late times it naturally transitions to a negative‐pressure dark energy–dominated phase, reproducing the observed cosmic acceleration \citep{MCG1, MCG2}. This smooth connection between the two cosmic epochs within a single framework provides a powerful mathematical tool and physical picture for exploring unified models of dark matter and dark energy \citep{Bamba2012}.
  
In addition to the scalar‐field formulation of the MCG model, we also develop a systematic method for obtaining a class of integrable  scalar-field cosmological models \citep{Wu}. Given a specified function $Y(X)$ and energy density $\rho(\phi$), we can derive the corresponding pressure $p(\phi)$, potential $V(\phi)$, as well as the equation of state $p(\rho)$. Using this method, we propose a new scalar-field ACG model, for which the resulting equation of state takes the form $p=\alpha\rho\ln\rho+\beta\rho$. The procedure is detailed in Appendix \ref{appendix:ingegrable}. 

While the Chaplygin-type models capture key features of cosmic acceleration \citep{Pourhassan2013, Kahya2015, Li2019, Aljaf2021}, their viability still remains under investigation and debate \citep{Carturan2003, Sandvik2004, Avelino2004, Fabris2011, Hashim2025}. To assess the feasibility of Chaplygin-type models as unified dark energy candidates and their consistency with current cosmological observations \citep{Riess2004, Bennett2013, Hinshaw2013, Bennett2014, Riess2021, Cunha2008, Farooq2017, Huang2020, Planck2020}, 
we conduct a full likelihood analysis of the four cosmological models, $\Lambda$CDM, GCG, MCG, and ACG, using the Pantheon+ SNeIa dataset with $0.001 < z < 2.26$ \citep{Brout2022, Scolnic2022}. We employed the MCMC technique, incorporating the full statistical and systematic covariance matrix together with Cepheid-calibrated distances. The analysis focused on four key cosmological parameters: the Hubble constant $H_0$, the present-day equation-of-state parameter $\omega_0$, the transition redshift $z^{\star}$, and the age of the Universe $t_0$. The posterior constraints of these cosmological parameters are summarized in Table \ref{fitting_table2}, while the corresponding marginalized posterior distributions are displayed in Figures~\ref{MCMC_LCDM_Color}--\ref{MCMC_ACG_Color}.

Our analysis of the Pantheon+ SNe Ia compilation calibrated with Cepheid distances shows that the inferred Hubble constant is highly consistent across all four models, clustering around $H_0 \simeq 72.55$–$72.83$ km s$^{-1}$ Mpc$^{-1}$. This consistency reflects that the local $H_0$ determination is insensitive to the assumed dark energy model once Cepheid calibration is imposed, as noted in previous studies \citep{Dhawan2020, Scolnic2018, Macaulay2019, Brout2022}. Furthermore, these values are in excellent agreement with other local distance-ladder measurements, including the SH0ES determination of $H_0 = 73.04 \pm 1.04$ km s$^{-1}$ Mpc$^{-1}$ \citep{Riess2022}, the Mira-calibrated SNeIa result of $H_0 = 73.3 \pm 4.0$ km s$^{-1}$ Mpc$^{-1}$ \citep{Huang2020}, and the H0LiCOW time-delay lensing measurement of $H_0 = 73.3^{+1.7}_{-1.8}$ km s$^{-1}$ Mpc$^{-1}$ \citep{Wong2020}, while remaining in $> 5\sigma$ tension with the lower CMB-inferred value from Planck ($67.4 \pm 0.5$ km s$^{-1}$ Mpc$^{-1}$; \citealt{Planck2020}). 

For $\Lambda$CDM, we obtain $\Omega_{m,0} \simeq 0.36^{+0.02}_{-0.02}$, generally consistent with the Pantheon+ SNe-only analysis ($0.334 \pm 0.018$; \citealt{Brout2022}) but slightly higher than the Planck 2018 baseline ($0.315 \pm 0.007$; \citealt{Planck2020}). The corresponding effective present-day equation of state is $w_{0} \simeq -0.64^{+0.02}_{-0.02}$, in comparison with $w_0 \approx -0.666$ from Pantheon+ \citep{Brout2022} and $w_0 \approx -0.685$ from Planck \citep{Planck2020}. In contrast, Chaplygin-type models replace $\Omega_{m, 0}$ with phenomenological parameters ($A^\star$ or $A_0$, $\alpha$, $\beta$) that encode the mixture of matter-like and dark-energy–like behavior \citep{Chaplygin4, Gorini2003}.

The transition redshift differs across models: $\Lambda$CDM gives $z^\star \simeq 0.52^{+0.04}_{-0.04}$, generally consistent with the widely supported range of $z^\star$ 
\citep{Riess2004, Wu2007b, Cunha2008, Farooq2017}, while Chaplygin-type models predict earlier acceleration with $z^\star \sim 0.85–0.92$. This reflects the unified dark fluid dynamics, in which the negative-pressure component declines more slowly than in $\Lambda$CDM and becomes important earlier \citep{Kamenshchik2001, Fabris2002, Gorini2003}. The inferred cosmic ages span $\sim 12.09–13.57$ Gyr, with $\Lambda$CDM and ACG slightly younger than the CMB-based values ($13.80 \pm 0.02$ Gyr from Planck; \citealt{Planck2020}), but consistent with independent lower bounds from stellar ages \citep{Bond2013, Tang2021, Valcin2021, Ying2023}. The longer age predicted by MCG illustrates the additional flexibility permitted by its two free parameters. Cosmic ages inferred from the oldest globular clusters can offer an independent cross-check \citep{Ying2023}.

The Chaplygin-type parameters we obtain are broadly consistent with earlier studies. For GCG, $A^\star \simeq 0.58^{+0.04}_{-0.03}$ with mildly negative $\beta$ agrees with constraints in \citet{Lu2009}, while the MCG parameters remain weakly constrained due to degeneracies noted in prior work \citep{MCG1, Lu2008, Paul2013}. By contrast, the ACG model gives tighter bounds ($\alpha \simeq 0.23^{+0.10}_{-0.08}$, $\beta \simeq -0.57^{+0.03}_{-0.03}$), indicating that SNe~Ia data can effectively limit its parameter space.

Overall, the consistency of Chaplygin-type models with the Pantheon+ likelihood analysis indicates that they can reproduce key cosmological observables while offering alternative explanations for cosmic acceleration \citep{Bamba2012, Wu}. Compared to $\Lambda$CDM, the Chaplygin-type models predict an earlier onset of cosmic acceleration and allow for a broader range of cosmic ages, illustrating their greater flexibility in late-time expansion histories. However, their added flexibility does not lead to significantly tighter constraints than $\Lambda$CDM, highlighting the importance of complementary datasets.

Future work will involve joint likelihood analyses combining SNe with BAO, CMB, and other external calibrators (e.g., TRGB) to break parameter degeneracies and extend the analysis to higher redshifts. Beyond background dynamics, a full calculation of the matter power spectrum using the MCMC-derived posterior constraints, together with direct comparisons to large-scale structure data, will provide a more stringent test of the perturbative viability of these models \citep{Avelino2004, Sandvik2004, Hashim2025}.

\acknowledgments

This research is supported by the NSTC grant 114-2112-M-004-001-MY2 from the National Science and Technology Council of Taiwan.\\

\emph{Software:} matplotlib \citep{Hunter2007}, numpy \citep{VanDerWalt2011, Harris2020}, scipy \citep{Virtanen2020}, corner \citep{Corner}

\newpage
\appendix
\vspace{0.5cm}
\section{Appendix A. scalar-field Model}
\label{appendix:ingegrable}

The new integrable scalar‐field model provides a systematic framework for constructing cosmological models derived from scalar field dynamics \citep{Wu}.

By specifying a function $\rho(\phi)$, we can systematically derive a complete set of cosmological quantities, including the pressure $p(\phi)$, potential $V(\phi)$, and the equation of state $p(\rho)$, thereby offering both theoretical clarity and observational applicability \citep{Panotopoulos2021}.\\

Starting from the Lagrangian density of a canonical scalar field $\phi$:
\begin{gather}
    {\cal L}(\phi)=-\frac{1}{2}g_{\mu\nu}\partial^{\mu}\phi\partial^{\nu}\phi-V(\phi),
\end{gather}

where $g_{\mu\nu} = \mathrm{diag}(-1,1,1,1)$ is the Minkowski metric and $\partial_\mu = \partial/\partial x^\mu$ denotes partial derivatives in four-dimensional spacetime. Applying Einstein's summation convention and the  principle of least action, the energy-momentum tensor is given by:
\begin{gather}
    T^{\mu\nu}=\partial^{\mu}\phi\partial^{\nu}\phi+g^{\mu\nu}{\cal L}(\phi),
\end{gather}

A perfect fluid is an idealized fluid with isotropic pressure in its local rest frame and no shear stresses or heat flux. The energy-momentum tensor for a perfect fluid is given by:
\begin{gather}
    T^{\mu\nu}=pg^{\mu\nu}+(p+\rho)u^{\mu}u^{\nu},
\end{gather}

where $u^{\mu}$ is the four-velocity of the observer in the comoving coordinate system. 
Assuming that matter is distributed in a homogeneous and isotropic universe~\citep{homogenousisotropic,homogenousisotropic2}, the scalar field has no spatial gradient, i.e., $\nabla\phi = 0$. 
From this, the energy density and pressure of the scalar field can be expressed as $\rho(\phi)=\dot{\phi}^2-{\cal L}(\phi)$ and $p(\phi)={\cal L}(\phi)$, respectively, leading to the following relations:
\begin{align}
     \rho(\phi)&=\frac{1}{2}\dot{\phi}^2+V(\phi)\label{eq3.4},\\
     p(\phi)&=\frac{1}{2}\dot{\phi}^2-V(\phi)\label{eq3.5},
\end{align}

where $V(\phi)$ is the potential and $\dot{\phi}$ is the time derivative of the scalar field. From Eqs.~(\ref{eq3.4}) and (\ref{eq3.5}), we get $\dot{\phi} = \sqrt{\rho + p} = \dot{\rho} d\phi/d\rho$. In the following, we consider only the positive root and omit the $\pm$ symbol. Using the Friedmann equation
\begin{gather}
    \rho=\frac{3H^2}{8\pi G}\label{ch1eq4},
\end{gather}

and adopting the convention \( \frac{8\pi G}{3} \equiv 1 \), we obtain \( H = \sqrt{\rho} \). Substituting this and $\dot{\rho} = \sqrt{\rho + p} \cdot d\rho/d\phi$ into the continuity equation:
\begin{gather}
    \dot{\rho}=-3H(\rho+p)\label{000},
\end{gather}

we derive the following nonlinear first-order differential equations:
\begin{align}
    p(\phi)&=\frac{1}{9\rho(\phi)}\left(\frac{d\rho}{d\phi}\right)^2-\rho(\phi)\label{ch3eq3.3},\\
    V(\phi)&=\rho(\phi)-\frac{1}{18\rho(\phi)}\left(\frac{d\rho}{d\phi}\right)^2\label{ch3eq3.4}.
\end{align}

Here, \( p(\phi) \) represents the pressure density and \( V(\phi) \) the potential, both originate from the scalar field \( \phi \). \\

Let $\rho(\phi) = \exp[X(\phi)]$ be the relation between the function 
$X(\phi)$ and the energy density $\rho(\phi)$.  

We define a new function
\begin{equation}
Y(X) = \left( \frac{dX}{d\phi} \right)^2 ,
\end{equation}
such that
\begin{equation}
X(\phi) = \int \sqrt{Y(X(\phi))} ~~ d \phi ~~ 
\end{equation}
Alternatively, \(\phi\) can be written as a function of \(X\):
\begin{equation}
\phi(X) = \int \frac{dX}{\sqrt{Y(X)}} .
\end{equation}

For those functions \(Y(\phi)\) with integrable 
\(\int dX/\sqrt{Y(X)}\), \(\phi\) can be written explicitly 
as a function \(\phi(\rho)\). As a result, the equation of state 
\(p(\rho)\) follows directly. For example, consider the case
\begin{equation}
\phi = \frac{2}{k_1}\sqrt{k_1 X + k_2} ,
\end{equation}
up to an integration constant, if
\begin{equation}
Y(X) =k_1 X + k_2,
\end{equation}
with constant parameters $k_1$ and $k_2$.  
Therefore,
\begin{equation}
X = \frac{k_1}{4}\phi^2 - \frac{k_2}{k_1} \equiv k\phi^2 - \lambda ,
\end{equation}
and 
\begin{equation}
\phi^2(\rho) = \frac{1}{k}\ln\rho + \frac{\lambda}{k}
\end{equation}
follows directly.\\

In summary, we obtain
\begin{eqnarray}
\rho(\phi) &=& \exp\!\left[k\phi^2 - \lambda\right], \label{eq:rho}\\[6pt]
\phi^2 &=& \frac{1}{k}\ln\rho + \frac{\lambda}{k}, \label{eq:phi}\\[6pt]
p &=& \left[\left(\frac{d\rho(\phi)/d\phi}{3\rho}\right)^2 - 1 \right]\rho
   = \left(\frac{4}{9}k^2\phi^2 - 1\right)\rho . \label{eq:p}
\end{eqnarray}

Hence, the equation of state can be written as
\begin{equation}
p = \alpha \rho \ln \rho + \beta \rho ,
\end{equation}
where
\begin{equation}
\alpha \equiv \frac{4}{9}k, 
\qquad 
\beta \equiv \frac{4}{9}k\lambda - 1 .
\end{equation}

\newpage

\end{CJK}

\newpage

\bibliographystyle{apj}
\bibliography{main}



\end{document}